\documentclass[12pt]{article}
\usepackage{graphicx}
\begin{document}
\centerline{\Large \bf The Krause-Hegselmann Consensus Model}
\centerline{\Large \bf with Discrete Opinions}

\bigskip
\bigskip

\centerline{Santo Fortunato}

\vskip0.3cm

\centerline{Fakult\"at f\"ur Physik, Universit\"at Bielefeld, 
D-33501 Bielefeld, Germany}

\noindent

\centerline{{\tt e-mail: fortunat@physik.uni-bielefeld.de}}

\bigskip

\begin{abstract}

The consensus model of Krause and Hegselmann can be naturally
extended to the case in which opinions are integer instead of 
real numbers. 
Our algorithm is much 
faster than the original version and thus more suitable
for applications.
For the case of a society 
in which everybody can talk to everybody else, we find that 
the chance to reach consensus is much higher as compared to
other models; 
if the number of possible opinions $Q\,{\leq}\,7$, in fact,
consensus is always reached, which might explain the stability 
of political coalitions with more than three or four parties.
For $Q>7$ the number $S$ of surviving
opinions is approximately the same independently of the size $N$ of the
population, as long as $Q<N$.
We considered as well the more realistic case
of a society structured like a Barab{\'a}si-Albert network; 
here the consensus threshold depends on the outdegree of the nodes
and we find a simple scaling law for $S$,
as observed for the discretized Deffuant model.

\end{abstract}

\bigskip


\bigskip

\section{Introduction}

Can statistical mechanics help to describe opinion dynamics?
The last few years have witnessed several attempts
in this direction \cite{Axel,Deff,HK,Sznajd,Galam,staufrev} and 
Monte Carlo simulations have become an important part of sociophysics 
\cite{weidlich}, enlarging the new field of interdisciplinary
applications of statistical physics \cite{librost}.
The starting point is represented by a random distribution of opinions,
which can be integer or real numbers, among a group of persons, or agents. 
Next, some simple dynamical mechanism is introduced so that, 
due to interactions between the agents,
the opinion of each agent changes over the time until, at some stage, 
a configuration is attained where the opinions are no longer modified by 
the dynamics and then remain the same. 
It is then interesting to study the possible stable opinion configurations
at the end of the process to see, for instance, if it is possible 
that all agents stick to one and the same opinion ("consensus") or whether
a polarization around fewer dominant opinions takes place. 
The most studied consensus models are those of Sznajd \cite{Sznajd}, Galam  
\cite{Galam}, Deffuant et al. \cite{Deff} and the one of Krause and Hegselmann
(KH) \cite{HK}.
The crucial differences between the models are the dynamical mechanisms 
to update the opinions, but they differ as well in other 
important aspects. 
For instance, in the original versions of the models of Deffuant and KH,
the opinion variable is a real number between 0 and 1, whereas in the models of Sznajd
and Galam it is an integer (this possibility was also 
studied for Deffuant in \cite{sousa}).
Besides, there are as well differences in the way
the topology of the society is conceived. In the Sznajd model the agents sit
on a lattice and can have opinion-affecting interactions only with their 
lattice neighbours; in the models of Deffuant and KH, instead, one assumes
a society in which each agent has the same probability to interact with
everybody, although recently scale free network topologies have also been considered
for Deffuant \cite{meyer,sousa}.
Among the above-mentioned consensus models, those of Sznajd and Deffuant 
are meanwhile quite well-known, as the convergence to the final configuration is 
relatively quick, which allows to simulate populations with millions of  
agents \cite{staufrev}. 

In this paper we focus instead on the KH model, which has not been 
investigated by many people so far. 
In its original version \cite{HK} one introduces
a real parameter $\epsilon$, called confidence bound. 
At every Monte Carlo step, the randomly selected agent $i$ with opinion $s_i$ takes 
the average of the opinions of those agents $j$ such that $|s_i-s_j|<\epsilon$.
This averaging process makes the algorithm very time-consuming compared, for
instance, to Deffuant, and is essentially the reason why
most people of the computational sociophysics community do not find it
attractive. 
However, the real interest behind consensus models is whether they are able to 
describe real situations, and it is not said that 
the faster the algorithms the better they are; 
the Sznajd model could effectively
simulate the distribution of votes among candidates in Brazilian 
and Indian elections
\cite{kertesz,gonzales}. 

We study here a 
modified version of the KH algorithm, where opinions take
integer values so that each individual has a finite number of possible
choices.
This is most often the case in real life; thinking for instance about
elections, the voters have a limited number of possible parties and/or
candidates among which to choose. We have used
several values for the number of opinions $Q$ and checked how many
different opinions survive in the final configuration.
Initially we have assumed a society in which every agent interacts with all 
the others. This is however quite an unrealistic situation; therefore 
we have as well checked what happens if the personal relationships 
within the society form a scale free network,
with few people having lots of friends and many having just a few.
To build the network we adopted the popular 
"rich get richer" strategy proposed by
Barab{\'a}si and Albert \cite{BA}. 

\bigskip

\section{The model}

Our opinions can take the values $1$, $2$, $3$, ...$Q$.
As far as the confidence bound $\epsilon$ is concerned, we will assume
in this paper that the agents are influenced only by the
individuals whose opinions differ
by at most one unit from theirs. This corresponds to the case $\epsilon=1/Q$
in the original KH model. Actually, in order to go smoothly 
to the continuous limit
one should introduce another parameter $L$, which is the
discrete confidence bound, i.e. the maximal distance between 
compatible opinions (for us $L=1$), and take the limit
$L,Q\rightarrow\infty$ by keeping $\epsilon=L/Q$ fixed. 

The algorithm starts by randomly distributing the opinions among
the agents. The use of integer-valued opinions spoils 
the original KH concept of "average of compatible
opinions", because such average in most cases would not be an integer.

\begin{figure}[hbt]
\begin{center}
\includegraphics[angle=-90,scale=0.5]{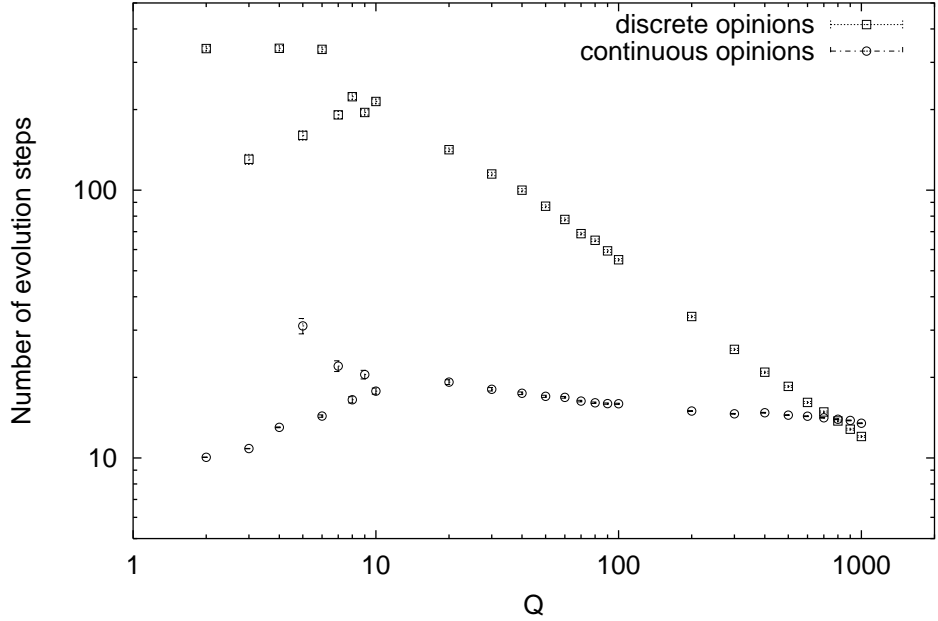}
\end{center}
\caption{\label{fig1}Number of evolution steps necessary to have 
convergence to the final configuration as a function of $Q$, for the standard 
KH model with continuous opinions and for our discretized version.
The number of agents is 1000 and for each value of
$Q$ we averaged over 1000 realizations.}
\end{figure}
\begin{figure}[hbt]
\begin{center}
\includegraphics[angle=-90,scale=0.5]{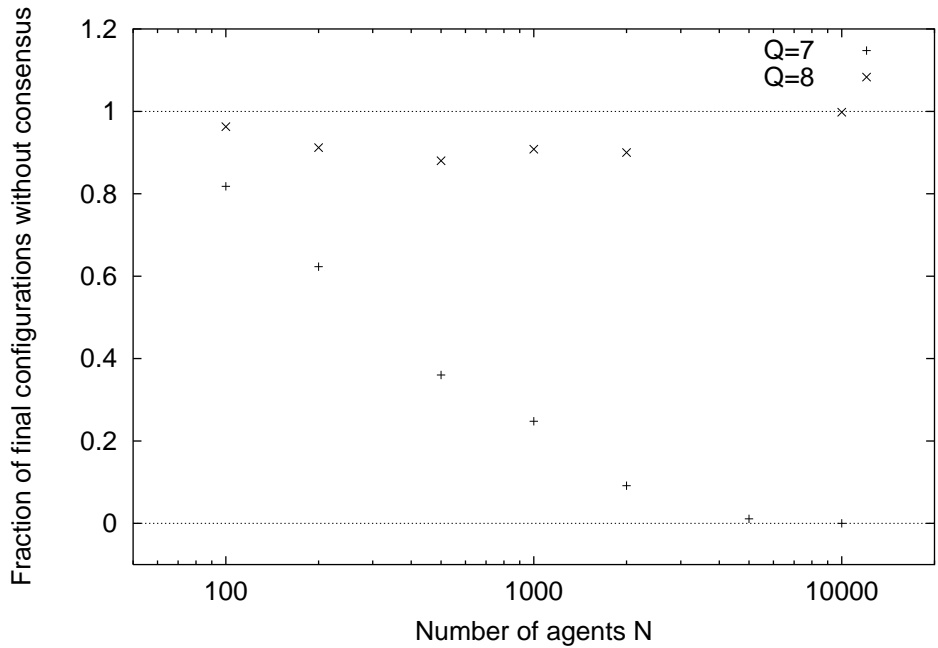}
\end{center}
\caption{\label{fig2}Fraction of final configurations in which 
no consensus is reached, as a function of the number $N$ of agents,
for $Q=7, 8$.} 
\end{figure}
\begin{figure}[hbt]
\begin{center}
\includegraphics[angle=-90,scale=0.5]{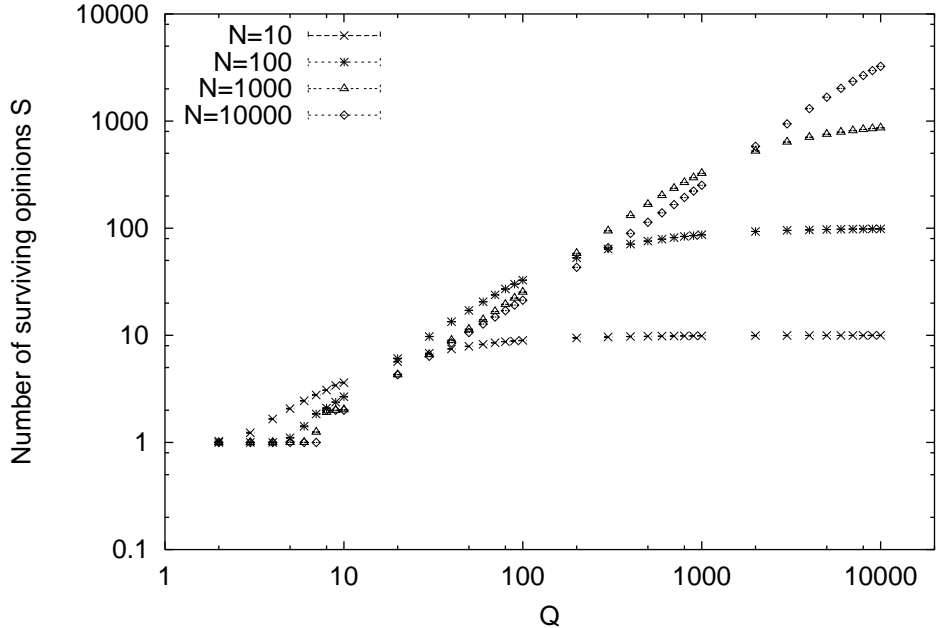}
\end{center}
\caption{\label{fig3}Number of surviving opinions $S$ as a function of $Q$
for a society where everybody interacts with everybody. 
We averaged over 1000 realizations.}
\end{figure}
\begin{figure}[hbt]
\begin{center}
\includegraphics[angle=-90,scale=0.5]{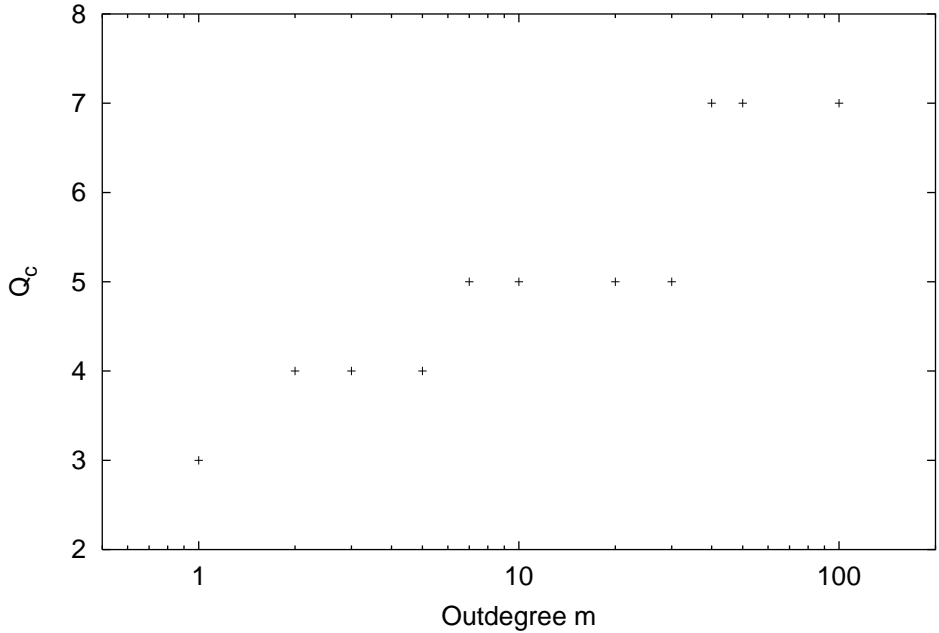}
\end{center}
\caption{\label{fig4}Threshold for complete consensus as a function 
of the outdegree $m$ of the network.}
\end{figure}
\begin{figure}[hbt]
\begin{center}
\includegraphics[angle=-90,scale=0.5]{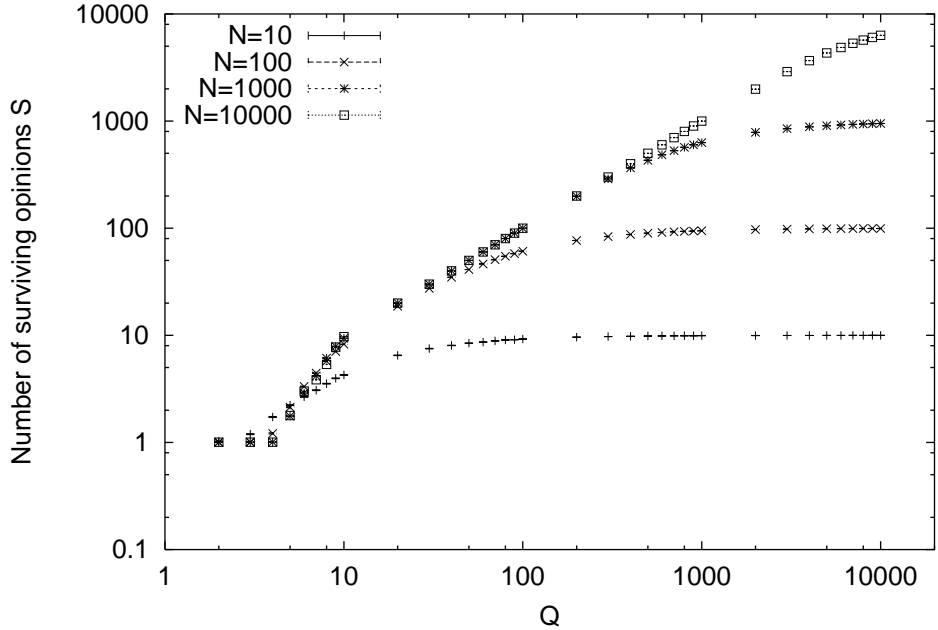}
\end{center}
\caption{\label{fig5}As Fig. \ref{fig3} but for a society 
where personal relationships are the edges of a Barab{\'a}si-Albert  
network. We averaged over 1000 realizations.}
\end{figure}
\begin{figure}[hbt]
\begin{center}
\includegraphics[angle=-90,scale=0.5]{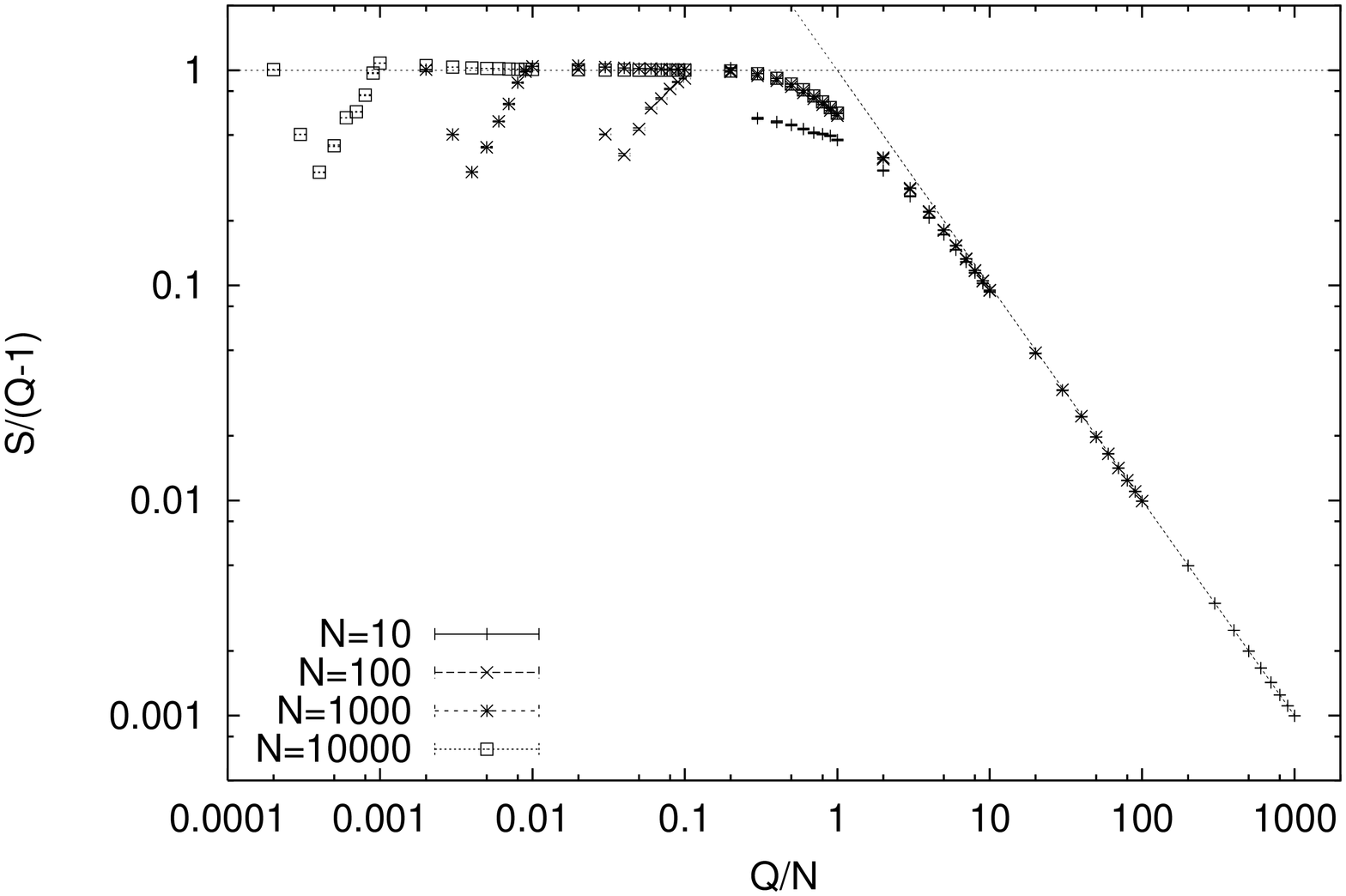}
\end{center}
\caption{\label{fig6}Scaled plot of the same data as in Fig. \ref{fig5}.
Apart from deviations for very small $Q$, where $S$ is the same 
independently of $N$, a reasonable scaling is observed. The two 
straight lines represent
the two situations where few opinions survive (horizonal) and
every agent keeps its own opinion (skew).}
\end{figure}

A possible way out is to re-interpret the spirit of the original
KH model in a probabilistic fashion. 
Suppose we want
to update the status of agent $i$, which 
has opinion $k$. The number of agents with compatible opinions are $n_{k-1}$,
$n_k$ and $n_{k+1}$ (respectively for opinions $k-1$, $k$ and $k+1$).
If the total number of compatible individuals is 
$n=n_{k-1}+n_{k}+n_{k+1}$, we say that agent $i$ takes 
opinion $k-1$, $k$ or $k+1$ with {\it probability}
$p_{k-1}=n_{k-1}/n$, $p_{k}=n_{k}/n$ and 
$p_{k+1}=n_{k+1}/n$, respectively. 
This is to our mind a natural extension of the KH model to discrete opinions,
and is the version we have used here. The status of the agents 
is updated sequentially, in an ordered sweep over the whole population; the
program stops if no agent changed opinion during an iteration.

The fact that the opinions are discretized allowed us to
speed up the algorithm compared to the continuous case.
In the latter the time to complete an iteration goes as $N^2$
($N$ is the size of the population),
because for each agent to update one needs to make a sweep over the whole
population to look for compatible individuals and calculate the 
average of their opinions. Here only the probabilities $p_k$ matter,
so we keep an array where the opinion histogram $n_k$
is stored ($k=1,2, .., Q$). When we update the agent $i$ with opinion $k$,
from the
histogram we derive directly the probabilities for the agent to 
take its next opinion. Suppose that agent $i$ takes opinion
$j$, what we need to do is to increase $n_j$ 
and to decrease $n_k$ by one unit, to get the new
opinion distribution, after that we can proceed to update a new agent.
In this way we avoid to count each time the number of
compatible individuals, which is
very time consuming, and the time needed to complete the iteration
goes as $(2L+1)N$ (for each agent one needs to make a sweep over
the $2L+1$ compatible opinion channels to get the probabilities, in our
case $2L+1=3$). With our algorithm systems with millions
of agents, unreachable by
standard KH, can be simulated (it took us
less than six hours to simulate one million agents on a PC).

We found that the convergence to a stable configuration
is much slower than in the original KH model. In Fig. \ref{fig1}
we plot the average number of sweeps necessary for convergence in the two cases,
as a function of $Q$ (we use the relation $\epsilon=1/Q$ to make a
correspondence between the two models). We see that, except for very
high values of $Q$, the number of evolution steps for our algorithm 
is an order of magnitude higher than for the continuous model.
This is most likely due to the stochastic character of our procedure; 
the opinion distributions vary more slowly if we allow jumps from one
opinion channel to the neighbouring ones 
with some probability
instead of sistematically shifting every agent to the average channel.
Moreover, the standard deviation of 
the average evolution time is much larger in the discrete than in the continuous
model, which hints to the presence of wild fluctuations. 

\bigskip

\section{Results}

\bigskip

The main result of our simulations is the existence of a threshold
$Q_c$, such that, for $Q\,{\leq}\,Q_c$, consensus is always reached. This is true for
both social topologies we have considered. For a society where each agent has
relationships with all others, we find that $Q_c=7$ as we can see in
Fig. \ref{fig2}. Here we plot the probability of having polarization
as a function of the number $N$ of agents, for $Q=7, 8$. By polarization
we mean that
more than just a single opinion survive in the final configuration.
The probability is given by the fraction of configurations
with polarization. 
For $Q=7$ this probability decreases
strongly with $N$ and for $N=10000$ all samples
presented a single final opinion. 
For $Q=8$, instead,
the probability for polarization is basically one 
for $N=10000$.
We remark that our threshold is higher than in the Sznajd model,
where $Q_c$ lies between $3$ and $4$ \cite{staufrev}, and in the discretized version of
Deffuant\footnote{In the original continuous versions the transition value
of the confidence bound for consensus is $\epsilon_c\,\sim\,0.4$ for Deffuant
and $\epsilon_c\,\sim\,0.21$ (our estimate) for KH.}, where $Q_c=2$.
This shows that the dynamics of the KH model is the most suitable to explain
how competing factions can find an agreement and to justify the stability 
of political coalitions with several parties like in Italy.

In Fig. \ref{fig3} we show how the number of surviving opinions $S$ 
varies with $Q$, for several $N$'s. 
We see that, as long
as $Q\,{\ll}\,N$, so that no finite size effects take over,
$S$ is approximately the same independently of the number of agents; 
the result holds for the Deffuant model as well \cite{bennaim}.

Let us now check what happens if we put the agents on 
a scale free network a la Barab{\'a}si-Albert. 
To build the network we must 
specify the outdegree $m$ of the nodes, i.e. the number of edges which originate
from a node. The procedure is dynamic; one starts from $m$ nodes which are all
connected to each other and adds further $N-m$ nodes one at a time.
When a new node is added, it selects $m$ of the preexisting nodes as neighbours,
so that 
the probability to get linked to a node is proportional
to the number of its neighbours. In all networks created in this way
the number of agents with {\it degree} $k$,
i.e. having $k$ neighbours, is proportional to $1/k^3$ for $k$ large,
independently of $m$. 

In our simulations we took the network as undirected, so that communication
between two neighbouring agents can take place in both directions.
We find again that there is a threshold $Q_c$ for the system
to evolve to complete consensus.
Interestingly, $Q_c$ depends on the outdegree $m$,
as shown in Fig. \ref{fig4}. 
In fact, the final stable configurations are those
in which each agent is surrounded only by agents which share its opinion
or are uncompatible, a solution of a special graph colouring problem; 
only in this case each agent will maintain
its opinion in the future with probability one.
If $m$ is small the average degree is small and
it is easier to reach such configurations even when 
just a few opinions are available.
Looking at Fig. \ref{fig4}, we see that $Q_c=3, 4$ 
for $m=1, 2$, respectively\footnote{As a matter of fact,
in the special case $m=1$
we find that consensus is not complete for $Q=3$, but is reached in 
about $80\%$ of the cases.}. 
No polarization is 
possible for $Q=2$ because, the network being connected, 
there would be at least two clusters sharing a border.
As each agent has at least $m$ neighbours
by construction, for large $m$ we expected 
to reach the threshold $Q_c=7$ that we have found in the case 
in which everybody is connected to everybody;
this is indeed true for $m$ larger than about 40.

The analysis of the number of surviving opinions $S$ is instead 
relatively independent of the outdegree $m$.
We chose $m=3$, in order to make comparisons with corresponding 
results for the discretized Deffuant model \cite{sousa}. 
Fig. \ref{fig5} shows $S$ as a function of $Q$ for different
population sizes. The pattern looks very similar to the one observed
in \cite{sousa} for Deffuant: for $Q$ not too small, $S$ equals $Q$
and only when $Q$ gets close to $N$, finite size effects take over
and $S$ converges towards $N$. In the latter case, the simulation
stops very early because most agents have
different opinions and therefore the chance 
for an agent to change its mind is small.
In \cite{sousa} a simple scaling behaviour 
of $S$ with $Q$ and $N$ was observed. The ansatz was 
\vspace{-0.5cm}
\begin{center}
\begin{equation}
S=(Q-1)f(Q/N); \hskip0.5cm f(x\rightarrow\,0)=1, f(x\rightarrow\,\infty)=1/x
\label{eq1}
\end{equation}
\end{center}
\vspace{0.2cm}

In Fig. \ref{fig6} we rescaled the data of Fig. \ref{fig5} according
to the ansatz of Eq. \ref{eq1}. The "teeth" below the horizontal
line refer to small values of $Q$, and we know that here $S$ equals 
one or is close to one,
independently of $N$; otherwise the scaling is good.

\bigskip

\section{Conclusions}

\bigskip

We have studied an extension of the Krause-Hegselmann consensus
model to integer-valued opinions, both when all
agents talk to each other and when they sit on the nodes of an undirected
 Barab{\'a}si-Albert
network. We assumed that only agents with opinions differing  
by at most one unit can influence each other.
A non-trivial implementation is necessarily probabilistic
and many more iterations are required for convergence compared to the 
standard model. On the other hand, 
our algorithm is much faster than the continuous version 
and therefore larger population sizes 
can be explored.
In a society where each agent can interact
with all others, when no more than seven
different opinions/positions are possible, the system
always evolves towards consensus if the size $N$ of the population
is large enough (of the order of $10^4$ agents or more).  
On the other hand, on a network-structured society, the threshold $Q_c$
depends on the minimal number of friends an agent can have; 
if this minimum is less than $30$, consensus is more difficult and
$Q_c$ can lower up to $3$. This versatility of the model 
makes it more suitable than others 
in order to explain how consensus can be reached in a variety of situations.
In Italy, for instance, the $80$'s were the years of the so-called
"pentapartito", a government's coalition of five parties. 
The stability of this coalition could be justified 
neither by the Sznajd model nor by Deffuant, but it is natural in
our model (although the concept of political stability in a country where 
fifty-seven governments alternated in fifty-eight years is questionable!).

In a fully connected society,
the number of final opinions $S$ is an intensive quantity, i.e.
independent of $N$ for large $N$. 

On a network
$S$ grows with $N$ if $Q$ and $N$ increase so that
the ratio $Q/N$ is constant, but is intensive if
$Q$ is kept fixed when $N\rightarrow\infty$.
When $Q$ is not too small $S$ is well described by the
same simple scaling function that reproduces the data
for the discretized Deffuant model \cite{sousa}. 
Moreover, we find that $S=Q$ for $Q$ 
of the order of ten or larger, so in a realistic society 
with more than ten different opinions/positions, the dynamics 
of the KH model is unable to suppress any of them.

\bigskip
\bigskip

{\bf \Large Acknowledgements}

\bigskip

I am indebted to D. Stauffer for introducing me into this fascinating field and
for many suggestions and comments.
I gratefully acknowledge the financial support of the DFG Forschergruppe
under grant FOR 339/2-1.

\end{document}